\documentclass[runningheads]{llncs}
\usepackage{graphicx}
\begin{document}
\title{Impact of Task Cycle Pattern on Project Success in Software Crowdsourcing}
%
%
\author{Razieh Saremi\inst{1} \and
Marzieh Lotfalian Saremi\inst{2}\and
Sanam Jena\inst{1}\and
Robert Anzalone\inst{1}\and Ahmed Bahabry\inst{1}}
\authorrunning{R. Saremi et al.}
%
\institute{Stevens Institute of Technology, Hoboken NJ, USA \and 
Concordia University, Montreal, Qc, Canada \\
\email{\{rsaremi,sjena,ranzalon,abahabry\}@stevens.edu,  m.lotfa@encs.concordia.ca }\\
}
\let\oldmaketitle\maketitle
\renewcommand{\maketitle}{\oldmaketitle\setcounter{footnote}{0}}

\maketitle              
\begin{abstract}


Crowdsourcing is becoming an accepted method of software development for different phases in the production lifecycle. Ideally, mass parallel production through Crowdsourcing could be an option for rapid acquisition in software engineering by leveraging infinite worker resource on the internet. It is important to understand the patterns and strategies of decomposing and uploading parallel tasks to maintain a stable worker supply as well as a satisfactory task completion rate.

This research report is an empirical analysis of the available tasks' lifecycle patterns in crowdsourcing. Following the waterfall model in Crowdsourced Software Development (CSD), this research identified four patterns for the sequence of task arrival  per project: 1) Prior Cycle, 2) Current Cycle, 3) Orbit Cycle, and 4) Fresh Cycle.

\keywords{Crowdsourcing  \and Topcoder \and Task Lifecycle \and Task Failure Ratio.}
\end{abstract}

\section{Introduction}

The interest in crowdsourcing software development(CSD) is rapidly increasing in both industry and academia. In crowdsourcing, jobs that were traditionally done in-house would be distributed among a large, distributed group of crowd workers\cite{howe2008crowdsourcing}. Software projects are using crowdsourcing methods in different phases  of software design and production\cite{surowiecki2005wisdom}. While understanding the

patterns and strategies of decomposing and scheduling crowdsourced are important to maintain stable worker supply as well as satisfactory task completion rate, much of existing studies are only focusing on individual task level, such as task pricing, task similarity, and task diversity \cite{faradani2011s}\cite{alelyani2017context}\cite{mejorado2020study}\cite{10.1007/978-3-030-50017-7_7}, and worker recommendation and behavior models\cite{latoza2015crowdsourcing}\cite{saremi2015dynamic}\cite{saremi2017leveraging}.

Crowd workers usually choose to register, work, and submit for certain tasks with the satisfactory monetary prize and comfortable level of dedicated effort, as monetary prize typically represents a degree of task complexity as well as required competition levels \cite{faradani2011s}\cite{archak2010money}.  Although sometimes Award is simply representing a specifically required skill to perform the task, it is one of the main factors influence crowd software workers in terms of the number of registrants and consequently, the number of submissions \cite{yang2015award}. Therefore, pricing tasks could be a huge challenge in decomposing the projects to mini-tasks and time of uploading them in the platform, yet by uploading more number parallel tasks, crowd workers would have more choice of utilized tasks to register for and consequently the chance of receiving more number of completed tasks is higher\cite{saremi2015empirical}. Since existing studies on general crowdsourcing reported limited or unpredictable results \cite{faradani2011s}\cite{archak2010money}\cite{latoza2015crowdsourcing}, it is a good opportunity to focus on parallelism on uploading the same project tasks.

For software managers, utilizing external unknown, uncontrollable, crowd workers would put their projects under greater uncertainty and risk compared with in-house development \cite{yang2015award}\cite{latoza2015crowdsourcing}. Understanding crowd worker’s sensitivity to the project stability and failure rate becomes extremely important for managers to make trade-offs among cost-saving, degree of competition, and expected quality in the deliverable\cite{yang2015award}.

To address these challenges, existing methods have explored various methods and techniques to bridge the information gap between demand and supply in crowdsourcing-based software development. These includes: studies towards developing better understanding of worker motivation and preferences in CSD \cite{faradani2011s}\cite{gordon1961general}\cite{difallah2016scheduling}\cite{yang2015award}, studies focusing on predicting task failure \cite{khanfor2017failure}\cite{khazankin2011qos}; studies employing modeling and simulation techniques to optimize CSD task execution processes\cite{saremi2018hybrid}\cite{saremi2019ant}\cite{urbaczek2020scheduling}; and studies for recommending the most suitable workers for tasks \cite{yang2016should} and developing methods to create crowdsourced teams\cite{wang2018solving} \cite{yue2015evolutionary}. However, there is a lack of study on the task execution flow in the field of software crowdsourcing.

To develop a better understanding of crowdsourcing-based software projects, this research reports an empirical study on analyzing Topcoder \footnote{ \url{https://www.topcoder.com/}}, largest software development crowdsourcing platform with an online community of 750000 Crowd Software workers. Topcoder intensively leverages crowdsourcing throughout the product implementation, testing, and assembly phases.

Topcoder started to explore crowdsourcing tasks in the form of competitions for software development, in which workers would independently create a solution and the winner will be chosen\cite{saremi2015dynamic}.

The remainder of this paper is structured as follows. Section II presents the background and review of available works. Section III outlines our research design. Section IV presents the empirical result. Section V discusses the findings and limitations and Section VI presents the conclusion and outlines several directions for future work.

\section{Research Background}

\subsection{Crowdsourced Platform Workflow}
A successful crowdsourcing platform contains three determinants: the characteristics of the project; the composition of the crowd; and the relationship among key players \cite{mao2017survey}. 
A systematic development process in a crowdsourcing platform starts from a requirements phase, where the project goals, task plan, and budget estimation are recognized. This will be performed through communication between the project manager, who may come from the crowd or the platform, and the requester, who pays for the solutions offered by the crowd. The outcome will be a set of requirements and specifications. These requirements are used as the input for the future architecture phase, while the application is decomposed into multiple components \cite{mingozzi1998exact}.
In CSD workflow , the task owner divides the project into many small tasks, prepares task descriptions, and distributes tasks through the platform. Each task is tagged with a pre-specified prize as the award \cite{mao2013pricing}\cite{yang2015award} to winners and a required schedule deadline to complete. On average, most of the tasks have a life span of 2-4 weeks from the first day of registration. 

\begin{figure}
\includegraphics[width=0.9\textwidth]{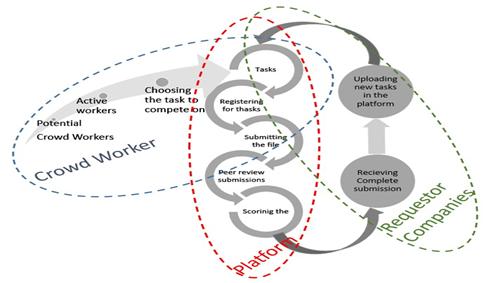}
\caption{Crowdsourcing Software Development Flows\cite{yang2015award}} \label{flow}
\end{figure}

Crowd software workers browse and register to work on selected tasks, and then submit the work products once completed. After workers submit the final submissions, the files will be evaluated by experts and experienced workers, through a peer review process, to check the code quality and/or document quality \cite{mao2017survey}. The number of submissions and the associated evaluated scores replicate the level of success in task success. In TopCoder, usually the award goes to the top 1 or 2 winners with the highest scores. If there are zero submissions, or no qualified submissions, the task will be treated as starved or cancelled. Figure \ref{flow} illustrates the CSD flow.

\subsection{Task Decomposition in Crowdsourcing}

In general, projects in crowdsourcing can be decomposed and executed in both independent and dependent tasks.  There are two methods of task decomposition: 1) horizontal task decomposition for independent subtasks and 2) vertical task decomposition for dependent sub-tasks\cite{jiang2014efficient}. In the horizontal decomposition method, workers dedicate efforts independently to their own subtasks for individual utility maximization, while in the vertical decomposition method, each subtask takes the output from the previous subtasks as input, and therefore the quality of each task is not only related to the worker’s effort but also associated with the quality of the previous tasks\cite{jiang2014efficient}.

Crowdsourcing complex tasks, i.e. software development tasks, generate heavy work-loads and require dedicated resources with high skill sets, which limit the pool of potential workers. In order to expand the qualified labor pool, it is essential to decompose software engineering tasks into smaller pieces.  However, software engineering tasks are often concerned with their specific contexts, for which decomposition may be complicated. This fact opens a discussion on different ways of decomposition based on a hierarchy of workflow \cite{stol2014two}.

The key factors of decompositions considered in this research are: smaller size of micro-tasks, larger parallelism, reducing time to market, and a higher probability of communication overhead.  The most common method in decomposing microtasks is asking individual workers to work on a task of specific artifact. This method will lead to some natural boundaries for software workers and there may be a need of defining new boundaries as well.

\subsection{Task Flow in Crowdsourcing}

Different characteristics of machine and human behavior create delays in product release\cite{ruhe2005art}. This phenomenon leads to a lack of systematic processes to balance the delivery of features with the available resources \cite{ruhe2005art}.
Therefore, improper scheduling would result in task starvation \cite{faradani2011s}. Parallelism in scheduling is a great method to create the chance of utilizing a greater pool of workers \cite{ngo2008optimized,saremi2015empirical}. Parallelism encourages workers to specialize and complete tasks in a shorter period. The method also promotes solutions that benefit the requester and can help researchers to clearly understand how workers decide to compete on a task and analyze the crowd workers performance \cite{faradani2011s}. Shorter schedule planning can be one of the most notable advantages of using CSD for managers \cite{lakhani2010topcoder}.

Batching tasks in similar groups is another effective method to reduce the complexity of tasks and it can dramatically reduce costs\cite{marcus2011human}. Batching crowdsourcing tasks would lead to a faster result than approaches which keep workers separate\cite{bernstein2011crowds}. There is a theoretical minimum batch size for every project according to the principles of product development flow \cite{reinertsen2009celeritas}. To some extent, the success of software crowdsourcing is associated with reduced batch size in small tasks.

\section{Empirical Study Design}

\subsection{Empirical Analysis}
\begin{figure}
\centering
\includegraphics[width=0.8\textwidth]{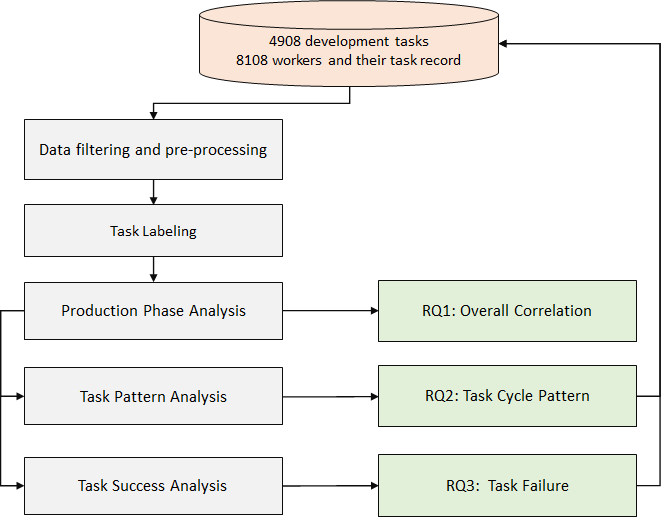}
\caption{Main flow of the proposed framework and relationship to research questions} \label{evaluation}
\end{figure}

To develop better understanding on the dynamic patterns in task supply and execution, we formulated research questions below  and investigated the different task cycle patterns in CSD and relationships among them in quantitative manners. The required steps to answer the the proposed research questions is illustrated in Fig \ref{evaluation}.

To investigate the evaluation framework, the following research questions were formulated and studied in this paper:

\begin{enumerate}

    \item \textit{RQ1 (Overall Correlation)}: How does different task production phases correlate with task success? 
    
     This research question aims at providing general overview of task success per software production phases in CSD platform;
    
    \item \textit{RQ2 (Task Cycle Pattern)}: Is there any task cycle pattern in CSD?
    
    Understanding task cycle patterns in CSD can be good measure to indicate project success;
    
    \item \textit{RQ3 (Task Failure)}: How does different task cycle patterns impact task Failure?
    
    The ratio of receiving not valid submission per task phase per task cycle pattern represents task success per identified task cycle.

\end{enumerate}

\subsection{Dataset}

The dataset from TopCoder contains 403 individual projects including 4,907 component development tasks (ended up with 4,770 after removing tasks with incomplete information) and 8,108 workers from January 2014 to February 2015 (14 months). Tasks are uploaded as competitions in the platform, where Crowd software workers would register and complete the challenges. On average, most of the tasks have a life cycle of one and half months from first day of registration to the submission’s deadline. When the workers submit the final files, it will be reviewed by experts to check the results and grant the scores.

\begin{table*}[!ht]
\caption{Summary of Metrics Definition} 
\centering 
\begin{tabular}{p{4cm} p{8cm}}
\hline
 Metrics & Definition \\ 
\hline
Task registration start date (TR) 	& The first day of task arrival in the platform and when workers can start registering for it. (mm/dd/yyyy) \\
Task submission end date (TS) & Deadline by which all workers who registered for task have to submit their final results. (mm/dd/yyyy) \\

Task registration end date (TRE)  & The last day that a task is available to be registered for.  (mm/dd/yyyy)\\
Monetary Prize (MP) & Monetary prize (USD) offered for completing the task and is found in task description. Range: (0, $\infty$).\\
Technology (Tech) & Required programming language to perform the task. Range: (0, \# Tech) \\ 
Platforms (PLT) & Number of platforms used in task. Range: (0, $\infty$). \\
Task Status & Completed or failed tasks \\
\# Registration (R)  & Number of registrants that sign up to compete in completing a task before registration deadline. Range: (0, $\infty$). \\
\# Submissions (S) & Number of submissions that a task receives before submission deadline. Range: (0, \# registrants]. \\
\# Valid Submissions (VS) & Number of submissions that a task receives by its submission deadline that have passed the peer review. Range: (0, \# registrants]. \\
\hline
\label{metrics}
\end{tabular}
\end{table*}

The dataset contains tasks attributes such as technology, platform, task description, monetary prize, days to submit, registration date, submission date, and the time-period (month) on which the task was launched in the platform.Then, we used expert based method and labeled associated phase with each tasks. The tasks attributes used in the analysis are presented in top section of Table \ref{metrics}.

Moreover, Topcoder clustered different task type to 7 groups as bellow:

\begin{enumerate}

    \item \textit{First2Finish}: The first person to submit passing entry wins
    \item \textit{Assembly Competition}: Assemble previous tasks
    \item \textit{Bug Hunt}: Find and fix available Bugs 
    \item \textit{Code}: Programming specific task
    \item \textit{UI Prototype}: User Interface prototyping is an analysis technique in which users are actively involved in the mocking-up of the UI for a system.
    \item \textit{Architecture}: This contest asks competitors to define the technical approach to implement the requirements. The output is a technical architecture document and finalized a plan for assembly contests.
     \item \textit{Test Suit}: Competitors produce automated test cases to validate the quality, accuracy, and performance of applications. The output is a suite of automated test cases.

\end{enumerate}

\section{Empirical Results}

\subsubsection{Overall Correlation (RQ1)}

\begin{figure}
\centering
\includegraphics[width=1\textwidth]{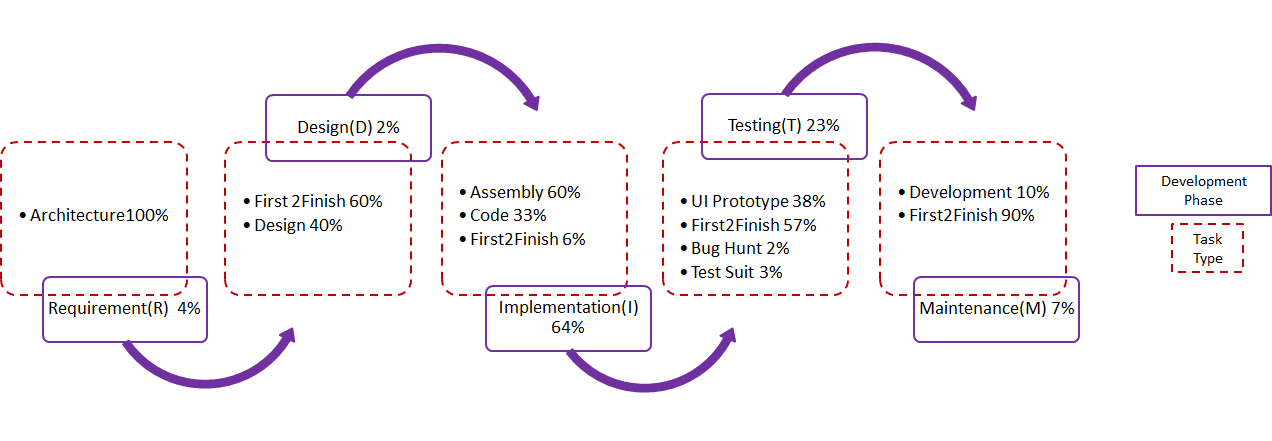}
\caption{Distribution of Task Failure Ratio among Different Development Phase in Task Lifecycle} \label{failure}
\end{figure}

It is reported that CSD platforms are following waterfall development model \cite{stol2014two}. Therefore, all tasks in the platform will follow development phases of Requirements, Design, Implementation, Testing, and Maintenance one after the other. 
Figure \ref{failure} presents the distribution of failure ratio for different task types per task phase in the task life cycle. 

As it is shown in figure \ref{failure}, highest task failure takes place in the implementation and testing phase with 64\% and 23\% respectively, while design and requirement are sharing less than 5\% of task failure each. One reason can be a lower number of the available task in these phases. Interestingly maintenance is holding 7\% of task failure in the platform. Moreover, First2Finish cluster contains tasks from all different development phase. This task type can be assigned to all different phases. As it is shown in figure\ref{failure}, 90\% of task failure in maintenance and 57\% of task failure in testing phase belong to this task type. While in implementation, only 6\% of task failure is under First2Finish task type, and interestingly 60\% of task failure happens in assembly tasks.

\subsubsection{Task Cycle Pattern (RQ2)}

\begin{figure}
\centering
\includegraphics[width=1\textwidth]{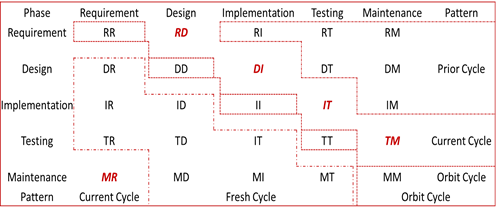}
\caption{Summary of Task Cycle in CSD Projects} \label{cycle}
\end{figure}

Waterfall development model in CSD makes each batch of tasks be always from a prior batch and a fresh batch. Following this sequence of task arrival we identified four cluster of task cycle patterns per project in CSD. In this study, the patterns are named as four clusters of Prior Cycle, Current Cycle, Orbit Cycle and fresh Cycle, which are defined as below:

Figure \ref{cycle} presents the summary of task cycles in a CSD platform. 

\begin{enumerate}

    \item \textit{Current Cycle} is the batch of tasks that are scheduled to complete following the initial task life cycle. 
    \item \textit{Prior Cycle } is the batch of tasks that went into task life cycle before the batch of current cycle arrives at the platform.
    \item \textit{Fresh cycle} is the batch of tasks that will start their life cycle after the current cycle arrives at the platform.
    \item \textit{Orbit Cycle} is the batch of tasks that all following the same development phase in the task life cycle.
   
\end{enumerate}

\subsubsection{Task Failure (RQ3)}

\begin{figure}
\centering
\includegraphics[width=1\textwidth]{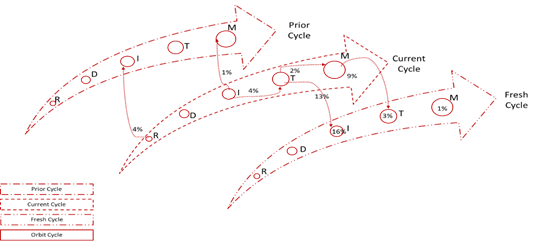}
\caption{Task Failure Pattern in Different Task Cycle} \label{success}
\end{figure}

Our analysis shows that on average 44\% of tasks in Topcoder are in Fresh Cycle while only 4\% of tasks are in the Prior cycle. Also, only 15\% of tasks are in Current Cycle and 37\% of tasks are in Orbit Cycle. 18\% and 20\% of task failure is respectively associated with Orbit Cycle and Fresh Cycle. Figure \ref{success} illustrates the details of task failure pattern in different task cycles.

\section{Discussion}

\subsection{Empirical Findings}

As it is shown in figure \ref{failure}, highest failure ratio happened in the same development phase. This can be the impact of task similarity. Our empirical result support that failure mostly happened due to receiving no submission. These observations raises the importance of studying both workers behavior and task similarity in details.

Moreover, identifying available task cycle in CSD revealed that the highest level of success is the result of task prior cycle while least level of task success is the result of fresh cycle task.  

\subsection{Threats to Validity}

First, the study only focuses on competitive CSD tasks on the TopCoder platform. Many more platforms do exist, and even though the results achieved are based on a comprehensive set of about 5,000 development tasks, the results cannot be claimed to be externally valid. There is no guarantee the same results would remain exactly the same in other CSD platforms.

Second, there are many different factors that may influence  task success, and task completion. Our task failure probability-focused approach is based on known task attributes in TopCoder. Different task failure probability-focused approaches may lead us to different, but similar results.

Third, the result is based on tasks description and completion only. Project level description and limitations are not considered in this research. In the future, we need to add this level of research to the existing one.

\section{Conclusion and future work}

To understand the probability of a tasks success in a crowdsource platform, one should understand the task cycle patterns in the platform. This research investigated available task cycle patterns in CSD.  Then analyzed tasks failure ratio in both software production phase and identified task cycle.

This study identified four patterns for sequence of task arrival  per project: 1) Prior Cycle, 2) Current Cycle, 3) Orbit Cycle and 4) Fresh Cycle. The empirical analysis support that the prior cycle led to the lowest task failure of 4\% while fresh cycle resulted n the highest level of task failure(i.e 44\%).

In future, we would like to use the identified task cycle and test them via crowdsourced task scheduling methods.

%
%
%
\bibliographystyle{splncs04}
%

\bibliography{name.bib}






\end{document}